\documentclass[12pt]{iopart}
\usepackage{epsfig}
\newcommand{\beq}{\begin{equation}}
\newcommand{\eeq}{\end{equation}}
\newcommand{\beqa}{\begin{eqnarray}}
\newcommand{\eeqa}{\end{eqnarray}}
\begin{document}
  \vspace*{-1.8cm}
  \begin{center}
  \epsfig{file=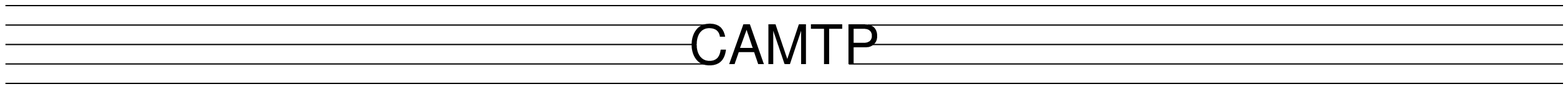,height=9mm,width=\textwidth}\vspace{2mm}
 Submitted to
 {\it Journal of Physics A}
  \hfill
   Preprint CAMTP/01-8\\
\hfill December 2001
   \end{center}\vspace{6mm}
\title{Semiclassical analysis of Wigner functions}
\author{Gregor Veble, Marko Robnik and Valery Romanovski}
\address{CAMTP Center for Applied Mathematics and Theoretical Physics,
University of Maribor, Krekova 2, SI-2000 Maribor, Slovenia}

\begin{abstract}
In this work we study the Wigner functions, 
which are the quantum analogues of the 
classical phase space density, and show how a full rigorous 
semiclassical scheme
for all orders of $\hbar$ can
be constructed for them without referring to the 
actual coordinate space wavefunctions from which
the Wigner functions are typically calculated. We find such a picture by a 
careful analysis around the stationary points of the main quantization 
equation, and apply this approach to
the harmonic oscillator solving it for all orders of 
$\hbar$.
\end{abstract}

\section{Introduction}

The Wigner functions (WFs) help us to picture the 
quantum states, that are typically represented as wavefunctions only in either 
configuration or momentum space, in the full phase space. They correspond
to the classical phase space density. According to the so-called Principle of 
Uniform Semiclassical Condensation (PUSC), 
they condense on a classical invariant
object (ergodic component) in the strict semiclassical
limit $\hbar\to 0$, when they become predominantly positive on this effective
support (Berry 1977b, Robnik 1998). 
They are 
of extreme importance when trying to compare and relate
the results of quantum mechanics
to the classical ones.

We typically obtain WFs by first finding the eigenstates in one of the usual
representations from which we then calculate the WFs as e.g. in equation
(\ref{eq:basic}). It is,
however, an intriguing question whether it is possible to handle the WFs
 as independent objects in the phase space without referring to the
corresponding eigenfunction. Such an approach was hinted at already by 
Heller (1976,1977). By a careful resummation of the Moyal bracket and 
a proper ansatz for the WF he managed to get an expression
of its time evolution. Interestingly enough, this result does not reduce
to a simple Liouville equation, the reason being the singular behaviour
of the WFs in the strict semiclassical limit. On the other 
hand, Berry (1977a) calculated the semiclassical approximation to the
WF by first using the semiclassical wavefunction, but since
the end result can be expressed in a way that does not put either the
coordinates or momenta into a privileged position, these approximations
to the WFs can be analyzed in the full phase space
independently from the wavefunction
approximations from which they were actually calculated. Ozorio de Almeida 
(1998) dealt with the Weyl representation in both classical and quantum 
mechanics, and managed to find a semiclassical periodic orbit formalism 
for the WFs that may be especially useful in the classically
chaotic systems.

Here we will try to find a quantum formalism that would expand the above
ideas in a way that would enable us to deal with WFs completely
independently from the eigenfunctions (in coordinate or momentum space),
 while at the same time we would like
to expand their semiclassical picture to all orders of $\hbar$. Osborn
in Molzahn (1995) did a similar expansion for the Weyl symbols of operators,
which are the generalizations of the WFs to operators other
than the density operator. They, however, require that the symbols are
regular in the semiclassical limit, which is not true for the WFs
that have an essential singularity in this limit. Still, with their
approach it is possible to find the phase space picture of the Heisenberg 
time evolution operator and act with that one on the irregular WF
to get its time evolution.

\section{The Wigner-Weyl formalism}

We can represent the WFs within the broader Weyl formalism by 
which operators are assigned symbols that are functions of the phase space
coordinates and momenta. The Weyl representation of an operator $\hat A$
is given by
\beq
A(q,p)=\int \left< q+x/2 \right|  \hat A \left| q-x/2 
\right> \exp\left(- i p x/ \hbar \right) dx.
\eeq
If $\hat A$ is self adjoint, the symbol $A(q,p)$ is real. Also, by integrating
over $p$ and then $q$ one can see that
\beq
{\rm Tr} \hat A = \int dq\ \left< q \right|  \hat A \left| q \right>=
\frac{1}{2 \pi \hbar} \int dq\ dp\ A(q,p). \label{eq:traceweyl}
\eeq
The WF by definition is just the Weyl symbol of the density operator
 $\hat \rho= \left| \psi \right> 
\left< \psi \right|$ divided by $2 \pi \hbar$,
\beq
W(q,p)=\frac{1}{2 \pi \hbar}\int \psi^\dagger (q-x/2) \psi(q+x/2) 
\exp\left(-i p x/ \hbar \right) dx. \label{eq:basic}
\eeq
It has a nice property that
\beq
\int dq\ dp\ W(q,p)=1, \label{eq:normaliz}
\eeq
which follows from (\ref{eq:traceweyl}), meaning that the WF is
properly normalized over the whole phase space. Normalization 
(\ref{eq:normaliz}) follows also clearly from (\ref{eq:basic})

Operators can be represented as being elements of a linear space.
We can find a basis and a scalar product within this space that will make
the manipulations of Weyl symbols easier. We can see that the trace of the
product of an operator with the adjoint of another operator,
\beq
\left<\hat A,\hat B\right>={\rm Tr}(\hat A \hat B^\dagger),
\eeq
indeed satisfies the conditions for it to be a scalar product of the two 
operators. This scalar product is real if the operators $\hat A$ and $\hat B$
are self adjoint.

A proper basis for our work is the family of operators
\beq
\hat \omega(q,p)=\frac{1}{\sqrt{2 \pi \hbar}}\int \left| q+x/2 \right> 
\left< q-x/2 \right|\  \exp\left(i p x/ \hbar \right) dx.
\eeq
By taking into account $\left< q| q^\prime\right>=
\delta(q-q^\prime)$ one can show that these operators are self adjoint, meaning
that
\beq
\fl \left<q_1\right|\hat \omega(q,p)\left|q_2\right> =  
\frac{1}{\sqrt{2 \pi \hbar}}\delta\left(\frac{q_1+q_2}{2}- q \right) 
\exp( i p (q_1-q_2)/\hbar) =
\left(\left<q_2\right|\hat \omega(q,p)\left|q_1\right> \right)^{\dagger},
\eeq
and therefore
\beq
\hat \omega(q,p)^\dagger=\hat \omega(q,p).
\eeq
Here $\delta(x)$ is the Dirac delta function.
These operators are also orthonormal with respect to the chosen scalar
product,
\beqa
\left<\hat \omega(q_1,p_1), \hat \omega(q_2,p_2)\right> =  \nonumber
\\
\frac{1}{2 \pi \hbar}\int \left< x\ |\ q_1+x_1/2\right>
\left<q_1-x_1/2\ |\ q_2+x_2/2\right>
\left<q_2-x_2/2\ |\ x\right>\cdot \nonumber
\\
 \cdot \exp\left(i (p_1 x_1+p_2 x_2)/\hbar\right) dx_1 dx_2 dx  \ =\
\delta(q_1-q_2) \delta(p_1-p_2). \label{eq:wigortho}
\eeqa
In deriving this relationship we only have to use the property
\beq
\int_{-\infty}^{\infty} \delta(x-a) \delta(x-b) dx=\delta(a-b)
\eeq
of the Dirac delta function.

With the help of the above expression the Weyl symbol of an operator
 $\hat A$ can be written as
\beq
A(q,p)={\sqrt{2 \pi \hbar}}\left<\hat A,\hat \omega(q,p)\right>.
\label{eq:aprod}
\eeq
Since the operators $\hat \omega(q,p)$ form a complete set of orthonormal
operators, we can also write
\beq
\hat A=\frac{1}{\sqrt{2 \pi \hbar}}\int A(q,p) \hat \omega(q,p) dq dp,
\eeq
which can be verified by insertion into (\ref{eq:aprod}).
This relationship is most helpful when one wants to find how the Weyl symbol
of the product of two operators can be expressed by their respective Weyl
symbols. Let
\beq
\hat C= \hat A \hat B.
\eeq
The Weyl symbol of the operator $\hat C$ is therefore
\beq
C(q_3,p_3)={\sqrt{2 \pi \hbar}}\left<\hat A \hat B,\hat 
\omega(q_3,p_3)\right>.
\eeq
By substituting
\beq
\hat A=\frac{1}{\sqrt{2 \pi \hbar}}\int A(q_1,p_1) 
\hat \omega(q_1,p_1) dq_1 dp_1
\eeq
and
\beq
\hat B=\frac{1}{\sqrt{2 \pi \hbar}}
\int B(q_2,p_2) \hat \omega(q_2,p_2) dq_2 dp_2,
\eeq
we obtain
\beq
\fl 
C(q_3,p_3)=\frac{1}{\sqrt{2 \pi \hbar}}\int A(q_1,p_1) B(q_2,p_2) 
{\rm Tr}\left(\hat \omega(q_1,p_1) \hat \omega(q_2,p_2) \hat 
\omega(q_3,p_3)\right)
 dq_1 dp_1 dq_2 dp_2.
\eeq
After a rather straightforward derivation not unlike
(\ref{eq:wigortho}) we obtain
\beqa
C(q_3,p_3)=\left( \frac{1}{ \pi \hbar} \right)^2 \int dq_1 dp_1 dq_2 dp_2 
A(q_1,p_1) B(q_2,p_2) \cdot \nonumber
\\
\cdot \exp\left(  2 i \left[p_1 (q_3-q_2)+p_2 (q_1-q_3)+p_3(q_2-q_1)
\right]/\hbar\right).\label{eq:weylprodukt}
\eeqa
The equation that determines the Weyl symbol of a product of two operators
is therefore an integral equation which makes it nonlocal. This will be
the main equation that will be dealt with in the following analysis of WFs.

\section{WKB expansion of Wigner functions}

We are now prepared to tackle the analysis of the WFs. We will
be dealing with the stationary problem of quantum mechanics, which in the 
standard picture leads to the search for eigenenergies and eigenstates of 
the Hamiltonian operator. In this standard picture, the main equation which
an eigenstate
$\left| \psi \right>$ must satisisfy is
\beq
\hat H \left| \psi \right>= E \left| \psi \right>.
\eeq
When dealing with WFs the core object we refer to is the
density operator which, for the case of a pure eigenstate, is written as
\beq
\hat \rho=\left| \psi \right>\left< \psi \right|.
\eeq
To ensure a proper solution, the quantization condition for the density 
operator in the nondegenerate case actually needs to satisfy a pair of 
equations (Curtright \etal 1998)
\beq
\hat H \hat \rho = E \hat \rho 
\eeq
and
\beq
\hat \rho \hat H = E \hat \rho.
\eeq
If we transform these equations to the Weyl formalism using 
(\ref{eq:weylprodukt}) we obtain the pair of equations
\beq
\left( \frac{1}{ \pi \hbar} \right)^2 \int dq_1 dp_1 dq_2 dp_2 \rho(q_1,p_1)
 H(q_2,p_2) 
\exp\left(\pm i \Delta_{123}/\hbar\right)=
E  \rho(q_3,p_3), \label{eq:wigmain}
\eeq
where
\beq
\Delta_{123}=2\left[q_1 (p_2-p_3)+q_2 (p_3-p_1)+q_3(p_1-p_2)\right],
\eeq
which corresponds to four times the area of a triangle spanned by the points
$(q_n,p_n)$, where $n=1\ldots 3$, in the phase space.

In a way similar to the usual WKB approach we can write the Weyl symbol of
the density operator as
\beq
\rho(q_n,p_n)=\exp(i \sigma_n/\hbar).
\eeq
The above may seem like a contradiction with the requirement that the WFs
need to be real. We will, however, see, that the above represents just a part
of the total solution and when all the parts are taken together the final
result can indeed be made real.
As in all the cases that follow, the index $n$ represents the evaluation of
the proper function in the point 
$(q_n,p_n)$. The equation (\ref{eq:wigmain}) then becomes
\beq
\left( \frac{1}{ \pi \hbar} \right)^2 \int dq_1 dp_1 dq_2 dp_2 
\exp(i \phi_{123}/\hbar)
H(q_2,p_2) =E, \label{eq:wigphase}
\eeq
where
\beq
\phi_{123}=\sigma_1-\sigma_3\pm\Delta_{123}.
\eeq

The approach to give us the main order $\hbar$ solution to the above problem
is the integration in the neighbourhood of the stationary points of the phase
 $\phi_{123}$. The equations that determine these points are
\beq
  \frac{\partial}{\partial x_n} \phi_{123}=0;\ \  x\in\{q,p\},n\in\{1,2\}.
\eeq
From this we determine the conditions for the stationary points
 $(q_1^{(0)},p_1^{(0)})$ and $(q_2^{(0)},p_2^{(0)})$, where 
$(q_3^{(0)},p_3^{(0)})$ is the point in which we wish to determine the 
WF, as being
\beqa
\frac{\partial \phi}{\partial q_1} & = &  \left(\frac{\partial \sigma^{(0)}}
{\partial q}\right)_1 
\pm 2 \left[p_2^{(0)}-p_3^{(0)}\right]  = 0, \label{eq:wigstatstart}
\\
\frac{\partial \phi}{\partial p_1}& = &  \left(\frac{\partial \sigma^{(0)}}
{\partial p}\right)_1 
\mp 2 \left [q_2^{(0)}-q_3^{(0)}\right]  =  0,
\\
\frac{\partial \phi}{\partial q_2}& = & 2\left[ -p_1^{(0)}+p_3^{(0)}\right] 
=  0,
\\
\frac{\partial \phi}{\partial p_2} &= & 2\left[ q_1^{(0)}-q_3^{(0)}\right] = 0
\label{eq:wigstatend},
\eeqa
where $\sigma^{(0)}$ denotes the lowest order $\hbar$ contribution to 
$\sigma$, as the basic stationary point analysis cannot reach any further.
The brackets $(\ldots)_i$ denote the function within them to be evaluated in
the corresponding stationary point
$\{q_i^{(0)},p_i^{(0)}\}$, where it is obvious that the points $1$ and
$3$ are the same. We can now shift our origin to the chosen stationary
point,
\beqa
\tilde q_1 & = & q_1-q_3^{(0)}, \label{eq:wigcoordstart}
\\
\tilde p_1 & = & p_1-p_3^{(0)},
\\
\tilde q_2 & = & q_2-q_3^{(0)} 
\mp \frac{1}{2}\left(\frac{\partial \sigma^{(0)}}
{\partial p}\right)_3,
\\
\tilde p_2 & = & p_2-p_3^{(0)}
\pm \frac{1}{2}\left(\frac{\partial \sigma^{(0)}}
{\partial q}\right)_3.
\label{eq:wigcoordend}
\eeqa
Rewriting equation (\ref{eq:wigphase}) into the new coordinates we obtain
\beq
\left( \frac{1}{ \pi \hbar} \right)^2 \int d\tilde q_1 d\tilde p_1 
d\tilde q_2 d\tilde p_2 
\exp\left(i \left[\tilde \sigma_1\pm 2(\tilde q_1 \tilde p_2-
\tilde q_2 \tilde p_1)\right]/\hbar\right)
\tilde H(\tilde q_2,\tilde p_2) =E, \label{eq:wigphasemod}
\eeq
where
\beq
\tilde \sigma_1=\sigma_1-\sigma_3-
\left[\left(\frac{\partial \sigma^{(0)}}{\partial q}\right)_1\tilde q_1+
\left(\frac{\partial \sigma^{(0)}}{\partial p}\right)_1\tilde p_1\right]. 
\label{eq:wigsigmatilde}
\eeq
and
\beq
\tilde H(\tilde q_2,\tilde p_2)=H(q_2, p_2). 
\eeq
For the quantities denoted by \~\ , the index $m$  naturally denotes evaluation
in the corresponding point $(\tilde q_m,\tilde p_m)$.

The analysis has so far been focused on the leading order $\hbar$ contribution.
We can use this leading order approximation to expand the analysis to all
orders in $\hbar$, with the leading order of this analysis being the same as
above, and we may write
\beq
\tilde\sigma_1=\sum_{n=0}^\infty (i \hbar)^n \tilde\sigma_1^{(n)}.
\eeq
We also perform a Taylor expansion to all orders in variables
$(\tilde q_n,\tilde p_n)$ for all quantities in equation
(\ref{eq:wigphasemod}), 
\beq
\tilde\sigma_1^{(n)}=\sum_{j=0}^{\infty} \tilde\sigma_1^{(n)(j)},
\eeq
\beq
\tilde H_2=\sum_{l=0}^{\infty} \tilde H_2^{(l)},
\eeq
where the indices $(j)$ and $(l)$ denote the order of the
homogeneous polynomials of the expansion with respect to the corresponding
coordinates 
$(\tilde q_n, \tilde p_n)$. These shifted coordinates are, just like in the
leading order analysis, obtained using the component 
$\tilde \sigma_1^{(0)}$ that represents the leading, zero order 
contribution in the expansion of $\tilde \sigma_1$ with respect to $\hbar$, as
given in equations
(\ref{eq:wigcoordstart})
-(\ref{eq:wigcoordend}). It will soon become apparent why such a choice of 
coordinates is proper.

By also taking into account
\beq
\exp(x)=\sum_{k=0}^\infty \frac{x^k}{k!},
\eeq
the equation (\ref{eq:wigphasemod}) can be written in a fully expanded form
\beqa
\left( \frac{1}{\pi \hbar} \right)^2 \int d\tilde q_1 d\tilde p_1 
d\tilde q_2 d\tilde p_2 \exp(\pm 2 i (\tilde q_1 \tilde p_2-
 \tilde q_2 \tilde p_1)/\hbar) \cdot \label{eq:wigexpanded}
\\
\sum_{l,k=0}^\infty \frac{1}{k!}\left[\frac{i}{\hbar}\sum_{n,j=0}^\infty
 (i \hbar)^n \tilde \sigma_1^{(n)(j)}
\right]^k \tilde H_2^{(l)}=E \nonumber
\eeqa

An important relationship to be used in the following analysis is
\beq
\int dq dp \exp (\pm 2 i q p/\hbar) q^m p^n= \pi \hbar \left (\pm i 
\frac{\hbar}{2}\right)^n
 n!\ \delta_{m,n}. \label{eq:wigintegral}
\eeq
It can be obtained by noting that
\beq
q^m \exp(\pm 2 i p q/\hbar)=\left (\mp i \frac{\hbar}{2}\right)^m 
\frac{\partial^m 
(\exp(\pm 2 i p q/\hbar))}
{\partial p^m}
\eeq 
and
\beq
\int dq \exp(\pm 2 i p q/\hbar)=\pi \hbar \delta(p)
\eeq
holds, and therefore we obtain
\beq
\int dq dp \exp (\pm 2 i q p/\hbar) q^m p^n=\pi \hbar \int dp\ p^n 
\left (\mp i \frac{\hbar}{2}\right)^m 
\frac{\partial^m}{\partial p^m} \delta(p).
\eeq
By an $m$-fold per-partes integration of the above expression 
we obtain the desired 
result
(\ref{eq:wigintegral}).

The equation (\ref{eq:wigexpanded}) shows that, upon integration, the factor 
$\delta_{m,n}$ in the expression (\ref{eq:wigintegral}) eliminates
all the contributions in the multiple sum of the expression
(\ref{eq:wigexpanded}) for which the order $l$ in the expansion of
 $\tilde H_2$ with respect to the homogeneous polynomials
does not match the polynomial order of the product formed by the various
Taylor expansion terms of the phase $\tilde \sigma_1$. These product terms
stem from the 
$k$-th order in the expansion of the exponential function and the subsequent
evaluation of the $k$-th power of the series that represents the full 
expansion of $\tilde \sigma_1$.
This leads to the condition
\beq
l=\sum_{p=1}^k j_p, \label{eq:wigmatch1}
\eeq
where $j_p$ represent the Taylor orders of those terms 
$\tilde \sigma_1^{(n)(j)}$ in the expansion of
$\tilde \sigma_1^{(n)}$ that form the chosen $k$-th order product term.

The main goal of this semiclassical analysis is to sort the various 
contributions of the equation
(\ref{eq:wigexpanded}) in terms of the orders of $\hbar$ with which they 
contribute. We denote the order of
$\hbar$ by which each term contributes to the total result by $o$. 
We again make use of the equation
(\ref{eq:wigintegral}). By carefully comparing it to the equation
(\ref{eq:wigexpanded}) we may see that for each contribution to the
multiple sum/product in equation
(\ref{eq:wigexpanded}) its appropriate order of $\hbar$ is given by
\beq
o= \frac{l}{2}+\sum_{p=1}^k \left[\frac{j_p}{2} +(n_p-1)\right].
\eeq
By also taking into account (\ref{eq:wigmatch1}), we obtain
\beq
o= -k + \sum_{p=1}^k \left[ j_p +n_p\right].\label{eq:wigmatch2}
\eeq
or, equivalently,
\beq
o= -k + l + \sum_{p=1}^k n_p.\label{eq:wigmatch3}
\eeq

It is very important to note that for each $\tilde \sigma_1^{(n_p)(j_p)}$ the
inequality 
\beq
j_p+n_p \geq 2
\eeq
holds. We can show this by first noting that $j_p\geq 1$ holds,
which follows from the fact that due to the construction (stationary point)
of
$\tilde \sigma$ the zero order Taylor contribution is equal to $0$.
Since, however, we are basing our analysis by an expansion around the
stationary point of the leading order, $n_p=0$, of the expansion of
$\tilde \sigma$ with respect to  $\hbar$, this means that for 
$n_p=0$, the linear, $j_p=1$, Taylor contribution is equal to $0$ as well. 

The above inequality leads to
\beq
k\leq o,
\eeq
which can be shown to hold true by 
\beq
 o=-k +\sum_{p=1}^k \left[ j_p +n_p\right]\geq -k +2k = k.
\eeq
This inequality tells us that the product terms 
that contribute with a given order $o$ in the $\hbar$ expansion 
can never comprise a greater number of factors $k$ 
than is the chosen $\hbar$ order $o$.
We can also show that
\beq
j_{p}\leq 2 o,
\eeq
holds, which can be seen by showing, from (\ref{eq:wigmatch2}),
\beq
j_{p}\leq \sum_{p=1}^k \left[ j_p +n_p\right]=o+k\leq 2 o.
\eeq
From this it follows that at a given order $o$ of the $\hbar$ expansion
the solutions can be sought locally as the order of the derivatives involved
can never be higher than the order $o$.
Another important inequality to consider is also
\beq
n_{p}\leq o, \label{eq:wigrelat}
\eeq
which follows from
\beq
n_{p}\leq \sum_{p=1}^k n_p=o+k-l \leq o,
\eeq
by also noting $k\leq \sum_{p=1}^k j_p=l$.
The relationship (\ref{eq:wigrelat}) tells us that only those terms
$\tilde \sigma_1^{(n)(j)}$ can contribute to a given order $o$ of the expansion
of the equation (\ref{eq:wigexpanded}) with respect to $\hbar$ for which
the order $n$ in the expansion of $\tilde \sigma$ does not exceed $o$.

All the above expressions lead to an important result that for each order
 $o$ of the expansion of the equation (\ref{eq:wigexpanded}) 
over $\hbar$ there is always a finite
number of terms involved. Even though the basic expansion could have been done
with respect to any point in the phase space, using the stationary point(s)
is the only choice which leads to the properties as given above. Using the
above properties the system becomes at least in principle locally solvable
since the equation for evaluating each order of the expansion of
$\sigma$ with respect to $\hbar$ contains only finite order derivatives
of the quantities involved. 

It is also important to observe that, at each order $o$ in the $\hbar$ 
expansion, the term with the highest order $n$ ($=o$) 
can only be linear ($k=1$)
and contains the first ($j=1$) derivative of
$\tilde \sigma_1^{(n)}$. This means that the gradient of 
$\tilde \sigma_1^{(n)}$
is, for each order $n$ in the expansion of $\tilde \sigma_1$ over the powers of
$\hbar$, determined by all
$\tilde \sigma_1^{(n^\prime)}$ for which $n^\prime<n$.

Using the above knowledge we may now try to rearrange the equation
(\ref{eq:wigexpanded}) and therefore (\ref{eq:wigphase}) with respect to
the orders of $\hbar$. As we determined above, for each order only a finite 
number of terms should contribute.
A properly reordered form of the equation (\ref{eq:wigphase}) is therefore
\beqa
\left( \frac{1}{\pi \hbar} \right)^2 \int d\tilde q_1 d\tilde p_1 
d\tilde q_2 d\tilde p_2 
 \exp(\pm 2 i (\tilde q_1 \tilde p_2-
\tilde q_2 \tilde p_1)/\hbar) \cdot \label{eq:wigsemiexpand} \\
\cdot \sum_{o=0}^{\infty}
\sum_{k=0}^{o} \sum_{l=0}^{2 o}\frac{1}{k!}\left(\frac{i}{\hbar}\right)^k
 \tilde H_2^{(l)} \sum_{\{ n_p,j_p \}_{k,l,o} }
\prod_{p=1}^k (i \hbar)^{n_p} 
\tilde \sigma_1^{(n_p)(j_p)}
 =E \nonumber, 
\eeqa
where we already dropped the terms that do not contribute upon integration
due to the relationships (\ref{eq:wigmatch1}) and (\ref{eq:wigmatch2}) not 
being fulfilled for them. The sum over
$\{ n_p,j_p \}_{k,l,o}$ is to be understood as a sum over all such
combinations of indices
$n_p$ in $j_p$ that, for given $k$, $l$ in $o$, match the conditions
(\ref{eq:wigmatch1}) and (\ref{eq:wigmatch2}).

With limiting the classical Hamiltonian to the form
\beq
H(q,p)=T(p)+V(q),
\eeq
which is by far the most common and for which the Weyl symbol
becomes equal to the classical Hamiltonian, the ordering
of terms with respect to the order of $\hbar$ becomes a little simpler
since all the mixed derivative contributions evaluate to zero in this case.
After integration only the terms where the products of derivatives of
$\tilde \sigma^{(n)}$ with respect to only $q$ or only $p$,
multiplied by the derivatives of  $\tilde H$ with respect to $p$ or $q$,
respectively, are preserved. In this case it becomes simpler to evaluate
the integration in the equation (\ref{eq:wigexpanded}) and the 
separations of contributions with respect to the order of $\hbar$ 
can be done in a semi-closed form.
The equation (\ref{eq:wigsemiexpand}) therefore becomes
\beqa
 \lefteqn{(H)_2 + \left( \frac{1}{\pi \hbar} \right)^2 \int d\tilde q_1 d\tilde p_1 
d\tilde q_2 d\tilde p_2 
 \exp(\pm 2 i (\tilde q_1 \tilde p_2-
\tilde q_2 \tilde p_1)/\hbar)\cdot}\\
& &\cdot \sum_{o=1}^{\infty}
\sum_{k=1}^{o} \sum_{l=1}^{2 o}\frac{1}{k!}\left(\frac{i}{\hbar}\right)^k
\left[  \tilde q_1^l \tilde p_2^l 
\frac{1}{l!} \left(\frac{\partial^l  H}{\partial p^l}
\right)_2 
\sum_{\{ n_p,j_p \}_{k,l,o} } \prod_{p=1}^k (i \hbar)^{n_p}
\left(\frac{\partial^{j_p}  \sigma^{(n_p)}}{\partial q^{j_p}}\right)_1
\frac{1}{j_p !} \right. \nonumber
+\\
& & + \left. \tilde p_1^l \tilde q_2^l 
\frac{1}{l!} \left(\frac{\partial^l  H}{\partial q^l}
\right)_2
\sum_{\{ n_p,j_p \}_{k,l,o} } \prod_{p=1}^k (i \hbar)^{n_p}
 \left( \frac{\partial^{j_p} \sigma^{(n_p)}}{\partial p^{j_p}} \right)_1
\frac{1}{j_p !} \right]  \nonumber 
 =E ,
\eeqa
where the powers of $p_1$ and $q_1$ that stem from the contributions
of $\tilde \sigma$ have already been joined and taken in front of the
product symbol. The brackets  $(\ldots)_i$ denote the function in them
to be evaluated at one of the the appropriate points
$\{q_i^{(0)},p_i^{(0)}\}$ which are given by 
(\ref{eq:wigstatstart}-\ref{eq:wigstatend}) that were obtained via the
stationarity condition using the lowest order  $\sigma^{(0)}$ of the
expansion of $\sigma$ with respect to the powers of $\hbar$.
The leading term in the $\hbar$ expansion, which is given simply by
 $ (H)_2$ is handled separately due to its somewhat different nature.

Using the equation (\ref{eq:wigintegral}) we obtain
\beqa
E =  (H)_2 + \sum_{o=1}^{\infty} (i \hbar)^{o}  \label{eq:wigurejen}
\sum_{k=1}^{o} (-1)^k\frac{1}{k!} \sum_{l=1}^{2 o}
\left(\pm \frac{1}{2}\right)^l
\\ 
\left[  \left( \frac{\partial^l  H}{\partial p^l}\right)_2 
\ \sum_{\{ n_p,j_p \}_{k,l,o} } \prod_{p=1}^k 
 \left( \frac{\partial^{j_p}  \sigma^{(n_p)}}{\partial q^{j_p}}\right)_1
\frac{1}{j_p !} \right. \nonumber
\\
+ \left. (-1)^l  \left(\frac{\partial^l  H}{\partial q^l}\right)_2  
\ \sum_{\{ n_p,j_p \}_{k,l,o} } \prod_{p=1}^k 
 \left( \frac{\partial^{j_p}  \sigma^{(n_p)}}{\partial p^{j_p}}\right)_1 
\frac{1}{j_p !}  \right] \nonumber,
\eeqa
which is the main equation to be solved, and is nicely sorted with respect
to the orders of $\hbar$.

\begin{figure}
\begin{center}
\epsfig{file=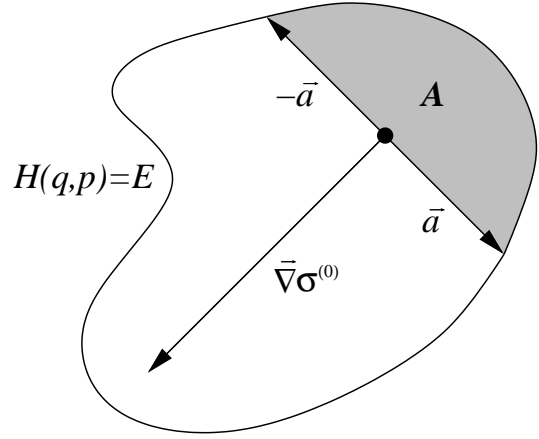,width=7cm}
\end{center}
\caption{Determining the lowest order $\hbar$ contribution to the 
phase of the WF. The closed curve represents the manifold (curve)
of constant energy. The gradient $\vec \nabla \sigma^{(0)}$ is orthogonal to
the vector $\vec a$, with the 
the actual value of the phase  $\sigma^{(0)}$ being given by the
area $A$ (see the main text for details).}
\label{fig:wignersketch}
\end{figure}

The lowest, zero order $\hbar$ contribution is an expression that looks
trivial at first, yet, however, it is quite involved,
\beq
 (H)_2=E,
\eeq
as it gives the pair of equations
\beq
H\left(q_1^{(0)}\pm \frac{1}{2}\left(\frac{\partial \sigma^{(0)}}
 {\partial p}\right)_1,\
p_1^{(0)}\mp \frac{1}{2}\left(\frac{\partial \sigma^{(0)}}{\partial q}\right)_1
\right)=E.
\label{eq:wigzeroorder}
\eeq
We can use this pair of equations to determine the local gradient of
the leading order phase contribution. It is determined by the chord
that can be spanned between two points on the curve (manifold) of the
constant energy, $H(q,p)=E$, for which $(q_1,p_1)$ is the center. The size
of the gradient is equal to the length of this chord, while at the same time
the gradient is orthogonal to it (see figure \ref{fig:wignersketch}).

Using this gradient we may also determine the actual value of the 
function $\sigma^{(0)}$. The easiest way to do this is to find all such chords
that are parallel to the one corresponding to the chosen point $(q_1,p_1)$
and lie between this point and the energy surface. The centers of these chords
form a path ${\bf s}$ in the phase space that starts on the energy surface
and ends in the point $(q_1,p_1)$. The change of phase along this path is given
by
\beq
\sigma^{(0)}= \int \vec \nabla \sigma^{(0)} \cdot d{\bf s}= 
\int l({\bf s}) dh=A,
\eeq
where $l$ denotes the length of the chord that corresponds to the
given point ${\bf s}$ and $dh$ is the component of $d{\bf s}$ that is
perpendicular to the chord. The value of this integral is $A$, which gives
exactly the area of the region
between the chord around a chosen point and the curve
of constant energy. This result is the same as obtained by Berry (1977a) where
the phase of the WF was determined using the leading
 semiclassical approximation for the WF.

The above equations typically give a pair of solutions. By properly connecting
these solutions at the caustics along with taking into account higher order
corrections leads to the quantization conditions and subsequent 
determination of the semiclassical energies (see Berry 1977a for details). 

Let us now consider the higher order $\hbar$ equations. All the terms that
contribute in the linear order of $\hbar$ give the pair of equations
\beqa
i \hbar \left[\mp \frac{1}{2} 
\left(
\left(\frac{\partial H}{\partial p}\right)_2
\left(\frac{\partial \sigma^{(1)}}{\partial q}\right)_1- 
\left(\frac{\partial H}{\partial q}\right)_2
\left(\frac{\partial \sigma^{(1)}}{\partial p}\right)_1
\right)\right.\nonumber
\\
-\left.\frac{1}{8}
\left(
\left(\frac{\partial^2 H}{\partial p^2}\right)_2
\left(\frac{\partial^2 \sigma^{(0)}}{\partial q^2}\right)_1+ 
\left(\frac{\partial^2 H}{\partial q^2}\right)_2
\left(\frac{\partial^2 \sigma^{(0)}}{\partial p^2}\right)_1
\right)\right]=0,
\eeqa
while the next order is already a more involved expression
\beqa
- \hbar^2 \left[
\mp \frac{1}{2} 
\left(
\left(\frac{\partial H}{\partial p}\right)_2
\left(\frac{\partial \sigma^{(2)}}{\partial q}\right)_1- 
\left(\frac{\partial H}{\partial q}\right)_2
\left(\frac{\partial \sigma^{(2)}}{\partial p}\right)_1
\right)\right.\label{eq:wig2ndorder}
\\
-\frac{1}{8}
\left(
\left(\frac{\partial^2 H}{\partial p^2}\right)_2
\left(\frac{\partial^2 \sigma^{(1)}}{\partial q^2}\right)_1+ 
\left(\frac{\partial^2 H}{\partial q^2}\right)_2
\left(\frac{\partial^2 \sigma^{(1)}}{\partial p^2}\right)_1
\right)\nonumber
\\
\mp \frac{1}{48}
 \left(
\left(\frac{\partial^3 H}{\partial p^3}\right)_2
\left(\frac{\partial^3 \sigma^{(0)}}{\partial q^3}\right)_1- 
\left(\frac{\partial^3 H}{\partial q^3}\right)_2
\left(\frac{\partial^3 \sigma^{(0)}}{\partial p^3}\right)_1
\right)
\nonumber
\\
+ \frac{1}{8}
\left(
\left(\frac{\partial^2 H}{\partial p^2}\right)_2
\left(\left(\frac{\partial \sigma^{(1)}}{\partial q}\right)_1\right)^2 
+\left(\frac{\partial^2 H}{\partial q^2}\right)_2
\left(\left(\frac{\partial \sigma^{(1)}}
{\partial p}\right)_1\right)^2 \right) \nonumber 
\\
\pm \frac{1}{32}
\left(
\left(\frac{\partial^3 H}{\partial p^3}\right)_2
\left(\frac{\partial^2 \sigma^{(1)}}{\partial q^2}\right)_1
\left(\frac{\partial \sigma^{(1)}}{\partial q}\right)_1
-\left(\frac{\partial^3 H}{\partial q^3}\right)_2
\left(\frac{\partial^2 \sigma^{(1)}}{\partial p^2}\right)_1
\left(\frac{\partial \sigma^{(1)}}{\partial p}\right)_1
\right) \nonumber 
\\
 +\left. \frac{1}{128}\left(
\left(\frac{\partial^4 H}{\partial p^4}\right)_2
\left(\left(\frac{\partial^2 \sigma^{(0)}}{\partial q^2}\right)_1\right)^2 
+\left(\frac{\partial^4 H}{\partial q^4}\right)_2
\left(\left(\frac{\partial^2 \sigma^{(0)}}
{\partial p^2}\right)_1\right)^2 \right)\right] \nonumber
=0.
\eeqa
Equations that correspond to higher orders are quite similar, and they
contain higher orders of derivatives of both $H$ and $\sigma^{(n)}$, 
while at the same time higher order products of $\sigma_1^{(n)}$ are involved.

\section{Harmonic oscillator}

As is almost customary in quantum mechanics, the test example for any 
new method is the harmonic oscillator. By properly scaling the coordinates
the Hamiltonian can be written as
\beq
H(q,p)=\frac{1}{2}\omega\left(p^2+q^2\right).
\eeq
As the Hamiltonian is quadratic in both the momentum and the coordinate,
only those terms of the equation
(\ref{eq:wigurejen}) can feature in its analysis that contain at most the
second order derivative of the Hamiltonian and, consequently, the phase
$\sigma$. At the same time the Hamiltonian is symmetric with respect to 
rotations around the phase space origin, and the same is true of the solutions
\beq
\sigma^{(n)}=\sigma^{(n)}(r)
\eeq 
which depend only on the distance
\beq
r=\sqrt{q^2+p^2}
\eeq
from the phase space origin.

Apart from the lowest order in the expansion over the powers of $\hbar$,
the proper equations for all orders $n\geq 1$ in the $\hbar$ expansion for this
system are given by 
\beq
\fl \frac{1}{2}\hbar^{n}\omega \left(\gamma(r) \frac{\partial \sigma^{(n)}}
{\partial r}\right)- \frac{1}{8}   \hbar^n\omega \left(\frac{\partial^2 
\sigma^{(n-1)}}{\partial r^2}+\frac{1}{r} \frac{\partial \sigma^{(n-1)}}
{\partial r}\right)+
\frac{1}{8}\hbar^n \omega \left(\sum_{m=1}^{n-1}\frac{\partial \sigma^{(m)}}
{\partial r}\frac{\partial \sigma^{(n-m)}}{\partial r}\right)=0, 
\label{eq:wigharmoneq}
\eeq
which is obtained from the pair of equations (\ref{eq:wigurejen}) 
by rewriting them in the radial coordinates, where the fact that the solution 
is symmetric with respect to rotations reduces both of these equations to the
expression above. We also introduced
\beq
\gamma(r)=\left(\frac{2E}{\omega}-r^2\right)^{\frac{1}{2}}.
\eeq
For the lowest order solution we use the equation (\ref{eq:wigzeroorder}) 
to obtain
\beq
\frac{\partial \sigma^{(0)}}{\partial r}=2 \gamma(r). \label{eq:wigh0}
\eeq
The next order in the expansion of $\sigma$ over powers of $\hbar$ is obtained
by the equation (\ref{eq:wigharmoneq}), which gives
\beq
\frac{\partial \sigma^{(1)}}{\partial r}=\frac{1}{2} 
\left( \frac{\gamma^{'}(r)}{\gamma(r)}+\frac{1}{r} \right), \label{eq:wigh1}
\eeq
and, after integration,
\beq
\sigma^{(1)}=\frac{1}{2}\left[\ln(\gamma(r))+\ln(r)\right] \label{eq:wigsig1}
\eeq

\begin{figure}[t]
\centerline{\epsfig{file=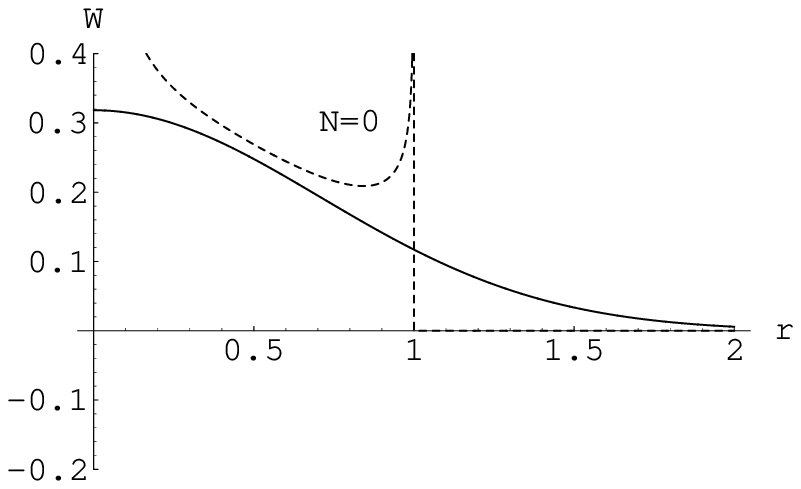, width=7cm}
            \epsfig{file=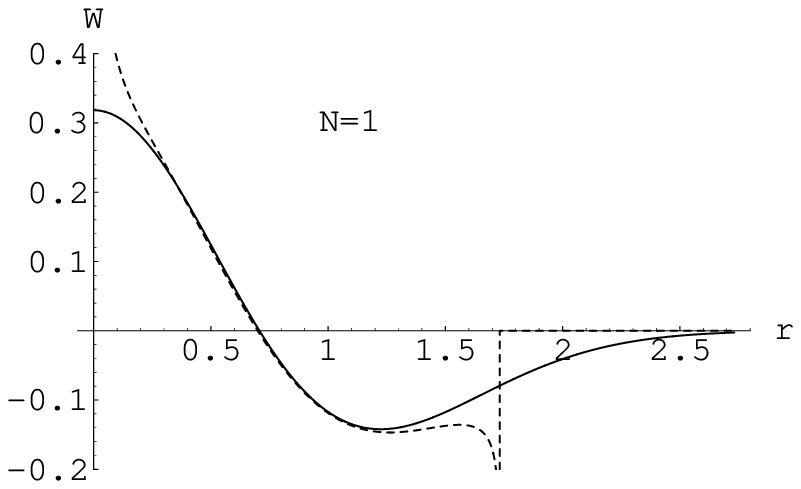, width=7cm}}
\centerline{\epsfig{file=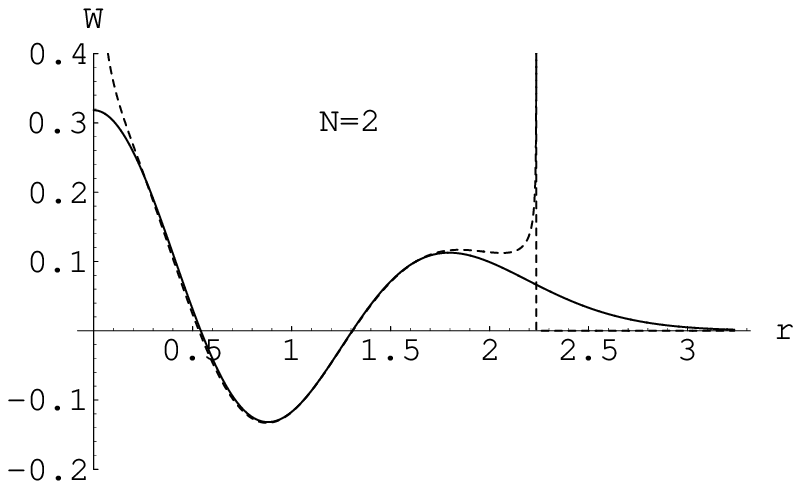, width=7cm}
            \epsfig{file=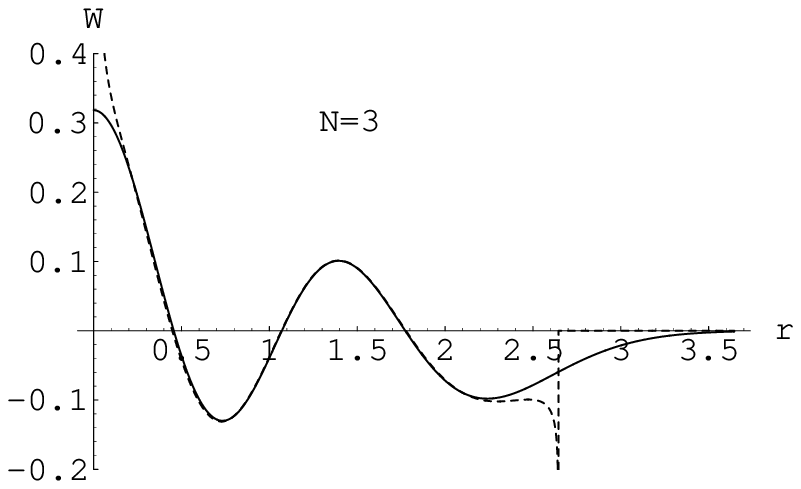, width=7cm}}
\centerline{\epsfig{file=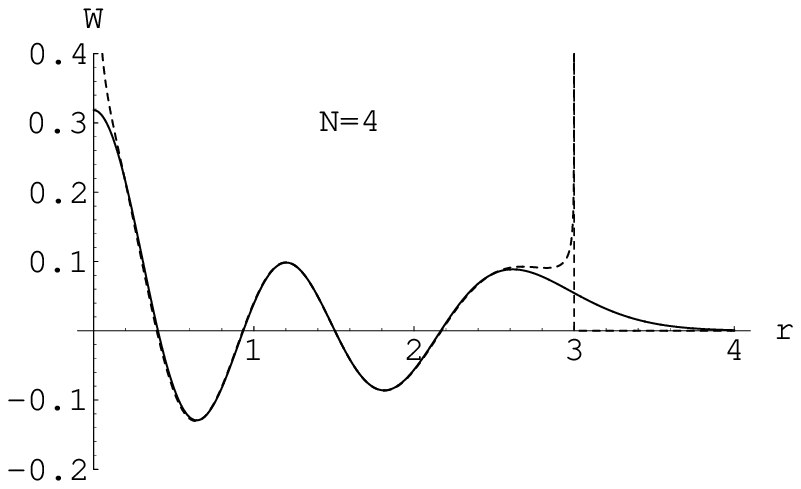, width=7cm}
            \epsfig{file=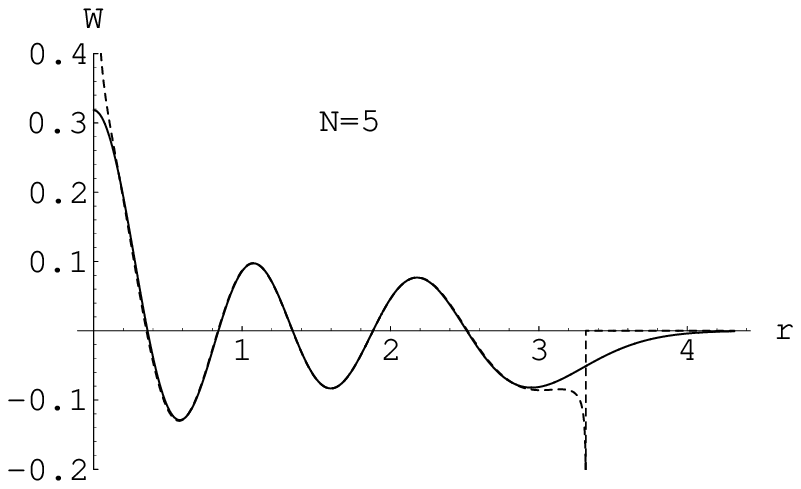, width=7cm}}
\centerline{\epsfig{file=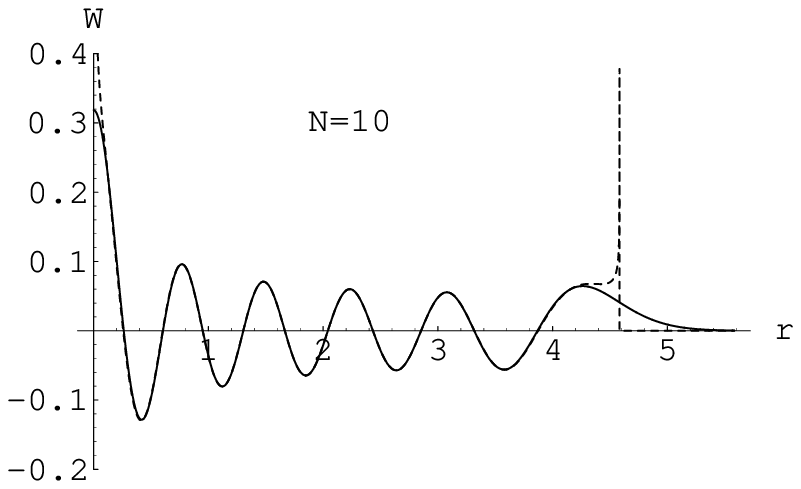, width=7cm}
            \epsfig{file=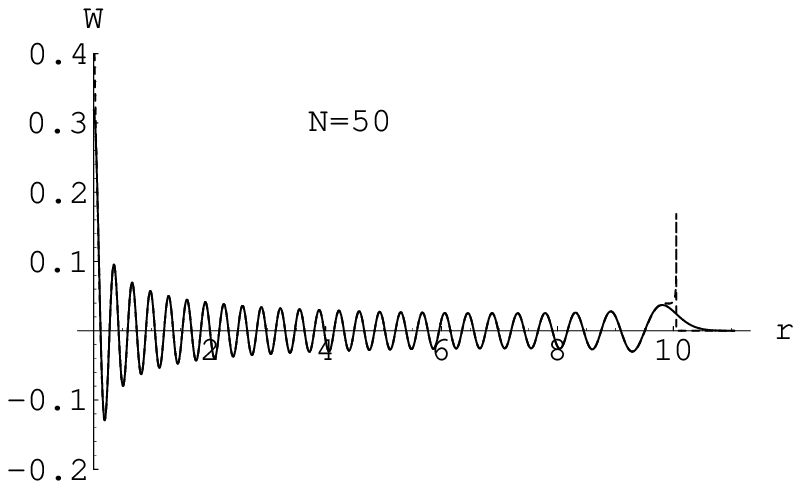, width=7cm}}
\caption{A comparison of the semiclassical (dashed) and the exact (full line)
WFs for various quantum numbers $N$, where we set $\hbar=\omega=1$. 
We used the first two 
terms in the $\hbar$ expansion of the phase to construct the semiclassical
solutions.}
\label{fig:wigharmon}
\end{figure}

In the figure \ref{fig:wigharmon} we show the semiclassical approximations
to the WFs (dashed) for various eigenstates 
by using the two contributions above along
with the exact solutions (full lines)
\beq
W_N(r)=\frac{1}{\pi \hbar}\exp\left(-\frac{r^2}{\hbar}\right)
L_N\left(\frac{2 r^2}{\hbar} \right)
\label{eq:wigexactform}
\eeq
where $L_N$ represents the Laguerre polynomial of order $N$.
We used the relationship
(Robnik 1998)
\beq
\int dq dp~W^2(q,p)=\frac{1}{2 \pi \hbar}. \label{eq:wignorm}
\eeq
to normalize the WFs.

\begin{figure}
\centerline{\epsfig{file=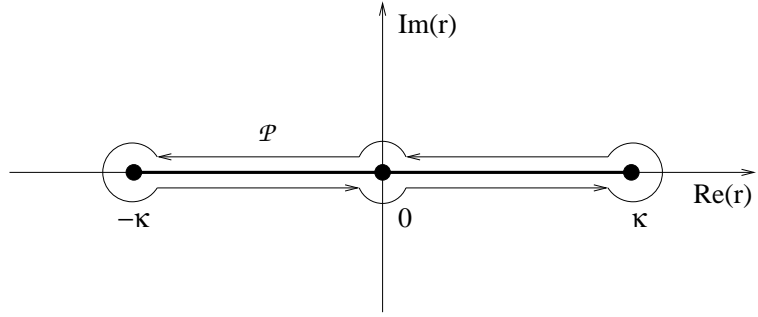, width=10cm}}
\caption{We show the complex plane of the variable $r$. With the thick
line we depict the cut in the complex plane that needs to be taken out of
the area of definition of the phase derivative, $\sigma^\prime$, in order
for it to be defined uniquely. We also show the contour ${\cal P}$ that goes
around the singularities in the points $r=\{0,\pm \kappa\}$, where 
$\kappa=\sqrt{2E/\omega}$, and from which the quantization condition is 
determined}
\label{fig:wigcut}
\end{figure}
To find these approximate solutions as well as to perform
 further analysis the 
solution in the whole complex plane needs to be carefully defined.
Due to the singularities of $\partial \sigma^{(0)}/\partial r$ 
and $\partial \sigma^{\prime(1)}/\partial r$ in the points 
$r=\{0,\pm \sqrt{2E/\omega}\}$ and the fractional power expressions in
both of them we can only make these derivatives uniquely defined on the
whole complex plane with the cut as shown in the figure
\ref{fig:wigcut}. This cut, on the other hand, just coincides with the main
domain on which we seek the solution. Therefore we obtain two contributions
on this cut that are the limits of the expressions as obtained by the limit of
approaching the cut from the upper or lower side, and therefore these 
expressions correspond to various sections of the path ${\cal P}$.
It is interesting to note that this cut is actually essential if we want the 
whole solution to be made real. Therefore we used the two branches when
constructing the total solution, which are obtained by taking the positive
and negative value of the square root in the definition of 
$\gamma(r)$ which then makes the total result real. 
By using both contributions
we may write the approximation to our Wigner function as
\beq
W(r)=A \cos\left(\frac{\sigma^{(0)}}{\hbar}+\theta\right) \exp(-\sigma^{(1)}),
\label{eq:wigsemifun}
\eeq
where $A$ is a real constant. Evaluating the integral of the 
equation (\ref{eq:wignorm}) therefore gives
\beq
\frac{1}{2 \pi \hbar}=
\int dq dp~W^2(q,p)\approx \frac{A^2}{2} \int dq dp~\exp(-2\sigma^{(1)}),
\eeq
where the value of the square of the trigonometric function was replaced by
its average which can indeed be done in the semiclassical limit
$\hbar\to 0$ where this function is rapidly oscillating. 
In our case this gives
\beq
\int dq dp~\exp(-2\sigma^{(1)})=2 \pi \int_0^{\sqrt{\frac{2E}{\omega}}} 
r~dr~ \frac{1}{\gamma(r) r}=\pi^2
\eeq
and therefore 
\beq
A=\frac{1}{\sqrt{\pi^3\hbar }}.
\eeq
We still need to determine the phase shift $\theta$, which is altered 
every time we encounter a singularity of 
$\partial \sigma^{(1)}/\partial r$ when traversing the path  ${\cal P}$ 
as shown in the figure \ref{fig:wigcut}. Although the expression
(\ref{eq:wigsig1}) tells us that the weight of the logarithmic contribution 
(which are responsible for the phase shifts)
when traversing
 the point $0$ is twice as strong as that at the other singular points,
we also need to take into account that traversing the path ${\cal P}$ we only
do a half of the full enclosure of this singular point. Upon encountering
any singularity along the contour ${\cal P}$ 
we therefore need to shift the phase
by $-\pi/2$. 

If at the same time we demand that the total phase upon the full traversal of
the contour ${\cal P}$ must change by an integer multiple of $2 \pi$, 
namely $2 \pi M$, as
the WF, which is exponentially dependent upon this phase, must be
singlevalued, this leads to the quantization condition which will be given 
in full detail later. The difference is that we now only take into account
the two lowest contributions of the expansion of the phase with respect to
$\hbar$, although this already gives the exact result for the
eigenenergies in our example. For odd $M$ it can be shown
 that we obtain semiclassical approximations
for the WFs that are odd with respect to the reflection of $r$,
which, however, contradicts the initial observation that the WFs must be
invariant with respect to rotations around the phase space origin. For the
even solutions ($M=2N$), on the other hand, we find that the phase shift 
in the 
expression (\ref{eq:wigsemifun})
needs to be $\theta=-\pi/4$ for $r>0$ if $\sigma^{(0)}(r=0)=0$ is chosen. 
This yields the explicit expression of equation (\ref{eq:wigsemifun}),
\beqa
\fl
W_N(r)=\frac{(-1)^N}{\pi \hbar \sqrt{\pi 
y\sqrt{2N+1-y^2}}} \sin\left(y \sqrt{2N+1 - y^2} -\right.
\nonumber
\\
\left.
- (2N + 1)\arccos \left(\frac{y}{\sqrt{2N + 1}}\right)+\frac{3 \pi}{4}\right);
\ \ 
r=\sqrt{\hbar} y
\eeqa
for $0<y<\sqrt{2 N+1} $. 
This is exactly the result one would
obtain by approximating the exact solution (\ref{eq:wigexactform}) using
the large $N$ approximation for the expression 
$\exp(-y^2/2) L_N(y^2)$ as found in 
(Szeg\"o 1959).

This now completes our approximate treatment of the WFs for the harmonic
oscillator, namely the two lowest orders, and now we turn to the exact treatment of the energy spectrum by considering all orders.
By using a straightforward yet lengthy procedure of induction it is easy 
enough to show that the expression below, when inserted into
(\ref{eq:wigharmoneq}), gives the correct solution to the problem,
\beq
\frac{\partial \sigma^{(n)}}{\partial r} = \omega 
\frac{\sum_{p=0}^{n} \alpha_p^{(n)} E^{n-p} (\omega r^2)^p}{r^n 
\left(\frac{2 E}{\omega}-r^2\right)^{\frac{3n-1}{2}}}, \label{eq:sigmaterm}
\eeq
where $\alpha_p^{(n)}$ are unknown rational coefficients, except for
$n=0,1$, where they are fixed by (\ref{eq:wigh0},\ref{eq:wigh1}).

We may now try to calculate the spectrum. We obtain it by taking a 
certain energy $E$ in the above equations and then trying to find such a 
value $E$ so that the WF
\beq
\rho=\exp\left(i \frac{\sigma}{\hbar}\right) 
\eeq
is singlevalued. This does not necessarily mean that the value of
the phase ${\sigma}/{\hbar}$ needs to be singlevalued, as it
may change by an integer multiple of $2 \pi$ when traversing any closed
path in the complex plane without changing the value of $\rho$ after
such a traversal.

It is interesting to note that using the classical WKB method to determine 
the eigenfunctions we may not directly link the condition of singlevaluedness
to the condition of the solutions being square integrable. Yet the 
singlevaluedness condition yields the correct values for eigenenergies in
systems that are exactly quantum solvable. Therefore the question
can be posed whether the two conditions are equivalent or does this only
apply to the solvable systems that usually possess some other special 
properties like solvability by the factorization method (Infeld and Hull 1951)
or other (Cooper \etal 1995, Robnik and Salasnich 1997a, 1997b, Robnik and
Romanovski 2000a, 2000b, Romanovski and Robnik 2000). 
For most of them we may find
the appropriate quantum canonical transformations (Lahiri, Ghosh and Kar 1998,
Veble 2001).
As we will see, using the singlevaluedness condition yields the proper 
solution in our case as well.

Let us now choose the closed path ${\cal P}$ in the complex $r$ plane as given
in figure \ref{fig:wigcut}
that encloses all the singularities
of our problem. These are found in the points 
 $r=\{0,\pm \sqrt{\frac{2E}{\omega}}\}$. 
The change of phase along this path,
\beq
\Delta \varphi= \frac{1}{\hbar} \oint_{\cal P} \sigma^\prime dr,
\eeq
is given by the residuum of $\sigma^\prime$ at infinity. We obtain it
by rewriting equation (\ref{eq:sigmaterm}) as
\beq
\frac{\partial \sigma^{(n)}}{\partial r} = \omega 
\frac{\sum_{p=0}^{n} \alpha_p^{(n)} E^{n-p} (\omega r^2)^p}{r^n 
\left(-r^2\right)^{\frac{3n-1}{2}}} \left(1+\frac{3n-1}{2}\frac{2 E}
{\omega r^2} +\ldots\right).
\eeq
The leading term of such an asymptotic series is of the order
$r^{-2n+1}$, with all the other terms comprising a higher negative power of 
$r$. As the residuum is given by the prefactor to the term containing $r^{-1}$,
the above expression can have a nonzero residuum only for
 $n=0,1$. By taking into account
equations (\ref{eq:wigh0}) and (\ref{eq:wigh1}), the 
evaluation of these residua therefore yields
\beq
\Delta \varphi = {2 \pi}\left[\frac{2 E}{\hbar \omega}- 1\right].
\eeq
By specifying that the above change of phase needs to be an integer multiple
of $2\pi$ we obtain the quantization condition for the energy
\beq
E=\hbar \omega\left[\frac{M}{2}+\frac{1}{2}\right],
\eeq
where $M$ is a nonnegative number. These solutions, however, also contain
those that yield WFs that are odd with respect to reflection of
the $r$ coordinate. Since the proper solutions need to be invariant with
respect to rotations around the phase space origin, only the even solutions
are the proper ones. This leads to $M=2N$ and therefore
\beq
E=\hbar \omega \left[ N + \frac{1}{2}\right].
\eeq
By constructing the full semiclassical WFs we therefore solved
the problem of the harmonic oscillator to all orders of $\hbar$ without 
referring to the actual wavefunctions. A similar analysis for the infinite
potential well (1-dim box potential) is in progress (Veble 2001,2002)

\section{Summary and conclusion}

By devising a full semiclassical analysis of WFs we managed
to rewrite quantum mechanics, that is typically considered in either only 
the momentum or coordinate representation, into an independent
 full phase space formalism. We
 obtained the full semiclassical equations to all orders of $\hbar$ 
for these functions. This
enabled us to solve the problem of the harmonic oscillator as an example.

It is easy enough to generalize the equations themselves to more than one
dimension. The problems arise when trying to solve for the main order 
contribution, as the mere condition of the appropriate chords lying on the
energy surface yields infinitely many solutions. We need to take other
conditions such as the singlevaluedness of the WFs with 
respect to all traversals in the phase space into account, and this is
far from trivial to implement. Most likely such a procedure, if it is found,
can function well only in 
classically integrable systems, or possibly for the regular states in 
the mixed systems, as nonintegrability and the chaotic motion associated with
it break the ordered structure of the classical phase space which is most
likely necessary for the generalization of the above procedure to work. 
Finding the
extension of the approach to more than one degree of freedom is therefore
the main goal of the work to follow.

\section*{Acknowledgements}

This work was supported by the Ministry of Education, Science and Sport of
the Republic of Slovenia, and by the Nova Kreditna Banka Maribor.

\section*{References}
\begin{harvard}
\item 
Berry M 1977a 
{\it Philosophical Transactions of the Royal Society of London A},
{\bf 287} 237 
\item 
Berry M V 1977b, {\it J. Phys. A: Math. Gen.} {\bf 10} 2083
\item
Cooper F, Khare A and Sukhatme U 1995 {\it Phys. Rep.} {\bf 251} 267-385
\item
Curtright T, Fairlie D and Cosmas Z 1998 {\it Physical Review D} {\bf 5802}
5002
\item 
Heller E J 1976 {\it Jour. Chem. Phys.} {\bf 65} (4) 1289
\item 
Heller E J 1977 {\it Jour. Chem. Phys.} {\bf 67} 3339
\item 
Infeld L and Hull T E 1951 {\it Rev. Mod. Phys.} {\bf 23} 21-68
\item 
Lahiri A, Ghosh G in Kar T M 1998 {\it Physics Letters A} {\bf 238} 239-243
\item 
Osborn T A in Molzahn F H 1995 {\it Ann. Phys. - New York} {\bf 241} (1) 79
\item 
Ozorio de Almeida A M 1998 {\it Physics Reports} {\bf 295} 265
\item 
Robnik M 1998 {\it Nonlinear Phenomena in Complex Systems} (Minsk) {\bf 1} 1
\item
Robnik M and Romanovski V 2000a {\it J. Phys. A: Math. Gen.} {\bf 33} 5093-5104
\item 
Robnik M and Romanovski V 2000b {\it Prog. Theor. Phys. Suppl.} {\bf 139} 
399-403
\item 
Robnik M and Salasnich L 1997a {\it J. Phys. A: Math. Gen.}  {\bf 30} 
1711-1718 
\item 
Robnik M and Salasnich L 1997b {\it J. Phys. A: Math. Gen.}  {\bf 30} 
1719-1729 
\item 
Romanovski V and Robnik M 2000 {\it J. Phys. A: Math. Gen.} {\bf 33} 8549-8557
\item
Szeg\"o G 1959 {\it Orthogonal Polynomials} (American Mathematical Society, 
New York)
\item 
Veble G 2001 {\it Ph.D. Thesis} (CAMTP, Universities of Maribor and Ljubljana)
\item 
Veble G 2002 {\it In preparation}
\end{harvard}

\end{document}